\documentclass[superscriptaddress,secnumarabic,nobibnotes,aps,prd,showkeys,noshowpacs,onecolumn,11pt]{revtex4-2}
\usepackage{eurosym}
\usepackage{graphics}
\usepackage{graphicx}
\usepackage{color}
\usepackage{epsf}
\usepackage{bm}
\usepackage{amsmath,amssymb,amsfonts,mathrsfs,amsthm}
\usepackage{latexsym}
\usepackage{enumerate}
\usepackage{comment}
\usepackage[dvipsnames,svgnames,x11names,table]{xcolor}
\usepackage[colorlinks = true,
            linkcolor = Blue,
            urlcolor  = Blue,
            citecolor = Blue,
            anchorcolor = Blue]{hyperref}
\usepackage{epstopdf}

\newcommand{\be}{\begin{eqnarray}}
\newcommand{\ee}{\end{eqnarray}}
\newcommand{\ben}{\begin{equation*}}
\newcommand{\een}{\end{equation*}}
\newcommand{\bean}{\begin{eqnarray*}}
\newcommand{\eean}{\end{eqnarray*}}

\newcommand{\bsub}{\begin{subequations}}
\newcommand{\esub}{\end{subequations}}

\def\bal#1\eal{\begin{align}#1\end{align}}

\newcommand{\disfrac}[1][2]{\displaystyle\frac}

\begin{document}

\title{Relativistic stars in $f(Q)$-gravity: Exact analytic solution for the power-law case $f(Q) = Q + b \: Q^\nu$}

\author{Nikolaos Dimakis}
\email{nikolaos.dimakis@ufrontera.cl}
\affiliation{Departamento de Ciencias F\'{\i}sicas, Universidad de La Frontera, Casilla
54-D, 4811186 Temuco, Chile}
\author{Alex Giacomini}
\email{alexgiacomini@uach.cl}
\affiliation{Instituto de Ciencias F\'{\i}sicas y Matem\'{a}ticas, Universidad Austral de
Chile, Valdivia, Chile}
\author{Andronikos Paliathanasis}
\email{anpaliat@phys.uoa.gr}
\affiliation{Institute of Systems Science \& Department of Mathematics, Faculty of
Applied Sciences, Durban University of Technology, Durban 4000, South Africa}
\affiliation{National Institute for Theoretical and Computational Sciences (NITheCS),
South Africa.}
\author{Grigorios Panotopoulos}
\email{grigorios.panotopoulos@ufrontera.cl}
\affiliation{Departamento de Ciencias F\'{\i}sicas, Universidad de la Frontera, Casilla
54-D, 4811186 Temuco, Chile}

\begin{abstract}
We investigate static spherically symmetric spacetimes within the framework of symmetric teleparallel $f(Q)$ gravity in order to describe relativistic stars. We adopt a specific ansatz for the background geometry corresponding to a singularity-free spacetime. We obtain an expression for the connection, which allows the derivation of solutions for any $f(Q)$ theory in this context. Our approach aims to address a recurring error appearing in the literature, where even when a connection compatible with spherical symmetry is adopted, the field equation for the connection is systematically omitted and not checked if it is satisfied. For the stellar configuration, we concentrate on the power-law model $f(Q)=Q+\alpha Q_{0}\left( \frac{Q}{Q_{0}}\right) ^{\nu }$. The de Sitter-Schwarzschild geometry naturally emerges as an attractor beyond a certain radius, we thus utilize it as the external solution beyond the boundary of the star. We perform a detailed investigation of the physical characteristics of the interior solution, explicitly determining the mass function, analyzing the resulting gravitational fluid properties and deriving the angular and radial speed of sound. 
\end{abstract}

\maketitle

\section{Introduction}

The analysis of cosmological observations \cite%
{Teg,Kowal,Komatsu,suzuki11,ade18,cco1,yy,yy1} challenges General Relativity
(GR) as a gravitational theory on very large scales. Modifying the
Einstein-Hilbert action by introducing geometric scalars, opens new
directions in the explanation of gravitational phenomena \cite{df2,md1,md2}. The resulting
modified field equations aim to provide corrections to GR, although the
deviations from the classical predictions need to be suppressed at scales
relevant to laboratory experiments or solar system tests. The rapid progress in observational cosmology and multimessenger astrophysics, allows now
for precision tests of fundamental physics, both on the scale of the observable
Universe, and in the strong gravity regime of relativistic compact
objects.

Possible quantum corrections to GR are sometimes expressed through modifications of the
Einstein-Hilbert Action by the addition of higher-order curvature invariants \cite%
{hg1,hg2,hg3,hg4}. The Starobinsky model of inflation \cite{hg5}
belongs to this family of theories; an $R^{2}$ term is introduced in order to drive the dynamics that explain the
early-time acceleration of the universe. The general family of modified theories in which the Starobinsky model belongs to
is the $f\left( R\right) $-gravity \cite{fr}. These theories are also related, through conformal mappings, to scalar tensor theories like e.g. the Brans-Dicke gravity \cite{oh1}.

Recently, Symmetric Teleparallel Equivalent General Relativity (STEGR) \cite%
{Nester:1998mp} and its extensions \cite%
{Koivisto2,Koivisto3,Baha1,sc1,sc2,mf1,mf2,mf4} have drawn the attention of
the academic community. For an extended review, we refer the interested reader to \cite%
{rev1}. In STEGR, the manifold is equipped with a metric tensor and a
symmetric, flat connection. From these properties of the connection, only the
nonmetricity tensor $Q_{\kappa \mu \nu }$ survives, and the ensuing nonmetricity
scalar, $Q$, is the fundamental element of the Lagrangian density of the gravitational Action Integral. The $Q$ is defined so that its difference from the Ricci scalar, calculated with the Levi-Civita connection of the metric tensor, is just a
boundary term \cite{rev1}. As a result the Action Integral of STEGR leads to
gravitational equations equivalent to those of GR. However, this equivalence is
fragile, as it is not carried over when one considers Lagrangians with a nonlinear dependence on the relative scalars.

The $f\left( Q\right) $-theory of gravity, which has been
introduced recently and aims to describe theories with a nonlinear dependence on $Q$, has been widely applied to describe the cosmic acceleration, including dark energy. For relevant references see \cite%
{re2,re1,re3,re4,re7,re8,re9,re10,re11,re12,re13,re14,re17,re18,re19,re20} and
references therein. At this point, it is important to mention that, while $%
f\left( Q\right) $-gravity suffers from the appearance of ghosts, or strong
coupling in cosmological perturbations \cite{ppr1,ppr2}, it nevertheless serves as a case
study to understand the effects of the connection in gravity. The choice of
connection in $f\left( Q\right) $-gravity is not always unique, meaning that
different connections lead to different gravitational models.

There are several studies in the literature that investigate static, spherically
symmetric spacetimes in $f\left( Q\right) $-gravity. Exact solutions
describing black holes have been presented in 
\cite{bh1,bh1a,bh1b,bh2,bh2a,bh2b,Calza}, while black holes in nonmetricity scalar-tensor theories
were studied in \cite{bh2,bh3}. The latter, is a more general theory which
includes $f\left( Q\right) $-gravity as a limiting case. The setting is similar to that which relates
Brans-Dicke to $f\left( R\right) $-gravity. In \cite{bh3},
an interesting result was found: namely, there are no static spherically symmetric
black hole solutions in $f\left( Q\right) $-gravity when the two metric
coefficients, $g_{tt}$ and $g_{rr}$, are constrained by $g_{rr}g_{tt}=-1$.
Moreover, the use of the coincidence gauge for static, spherically symmetric spacetimes
was discussed in \cite{Laur1}, and possible constraints from solar system tests in the theory are discussed in \cite{sol}. Recently neutron stars solutions in $f(Q)$ theory have been presented in \cite{Heisenbergnew}, and a stellar solution in the context of scalar nonmetricity in \cite{oursnew}.


As has been mentioned in \cite{otherpaper}, the timeline of the investigation of the properties of relativistic and compact stars, initiated in 1916 with Schwarzschild's solution to Einstein's field equations; an analysis that has been fundamental to astrophysics. The solution may be interpreted as the gravitational field generated by a body characterized by spherical symmetry \cite{Schwarzschild:1916uq}. That work boosted subsequent studies on stellar modelling and internal composition of astronomical objects characterized by ultra-dense matter content. Under those extreme conditions that cannot reproduced on terrestrial experiments, the Newtonian description becomes inadequate, and thus a full general relativistic treatment is required. Thanks to their unique properties, relativistic stars comprise ideal cosmic laboratories capable of constraining new physics and alternative theories of gravity.

When studying stars the authors usually focus their attention on astrophysical objects made of matter viewed as a perfect fluid, where there is a unique pressure along all spatial dimensions, $p_r=p_\theta=p_\phi$. However, celestial bodies are not necessarily made of isotropic fluids. In fact, under certain conditions matter content of compact objects may become anisotropic. The review article of Ruderman \cite{aniso1} long time ago mentioned such a possibility: there the author makes for the first time the observation that in very dense media anisotropies can be generated by interactions between relativistic particles. A comprehensive study of fluid spheres made of anisotropic matter was presented in \cite{Cadogan:2024mcl} in Newtonian as well as relativistic gravity. In the context of Einstein's General Relativity, anisotropic fluids have been extensively studied, with particular emphasis on stellar modeling and their impact on structural properties of compact objects \cite{Herrera1,Herrera2}.

In spite of many successful predictions of GR and its mathematical elegance, in most of the cases of interest the study and analysis of realistic physical situations comprises a formidable task, due to the fact that Einstein's field equations are highly non-linear coupled partial differential equations. Therefore, finding exact analytic solutions to the field equations of Einstein's gravity has always been an interesting and challenging topic, keeping researchers busy for decades. For exact known solutions to Einstein's field equations, see \cite{ExactSol}. Nowadays there is a number of approaches that allow us to obtain exact analytic solutions to the gravitational field equations, such as the Minimal Geometric Deformation \cite{Ovalle:2017fgl}, the vanishing complexity factor formalism \cite{herrera} as well as the Karmarkar condition in gravity \cite{karmarkar}.


In this work, we focus on analyzing interior static spherically symmetric
solutions within the framework of $f\left( Q\right) $-gravity. To realize this investigation, we consider a specific ansatz for the geometry and examine
the physical properties of the gravitational fluid. We derive a solution for the connection, which provides a general setting that can be applied to any $f(Q)$ theory in the search of stellar solutions. We make use of the relevant solution to investigate stellar configurations in a power-law theory of the form $f(Q)=Q+\alpha \,Q_{0}\left( \frac{Q}{Q_{0}}\right) ^{\nu }$. It follows that the de
Sitter-Schwarzschild geometry acts as an attractor for this model, allowing
us to define a boundary for the star and determine the mass function. 

This work demonstrates the necessary steps to derive a consistent solution in the context of $f(Q)$ gravity, using a choice of a connection that is compatible with any $f(Q)$ theory and allows for the satisfaction of the connection equation without the generation of additional constraints on the metric function outside of the metric equations. The primary objective of the article is to establish the mathematical and conceptual consistency of stellar solutions in $f(Q)$ gravity, rather than to propose a definitive astrophysical model. For this reason, in the example we present, we opt for the derivation of an exact solution by imposing a function $f(Q)$ of a power-law form and adopting a specific ansatz for the metric.

The structure of the paper is as follows: In Section \ref{sec2}, we briefly introduce symmetric teleparallel gravity
and its extension, the $f\left( Q\right) $-theory of gravity. Emphasis is
given to the limit in which General Relativity is recovered within $f\left(
Q\right) $-theory. In Section \ref{sec3}, we focus on static spherically
symmetric spacetimes. We introduce the symmetric and flat connection, which
inherits the four symmetries of the background geometry, and present the
gravitational field equations. We also discuss the de Sitter--Schwarzschild
limit to be used for the exterior solution. In Section \ref{sec4}, we introduce the model under consideration; a power-law $f(Q)$-theory which offers a reasonable deviation from GR. Observational cosmological constraints on this theory have been studied in \cite{power1,power2}. We adopt a specific ansatz for the
spacetime structure and analyze the evolution of the physical parameters of
the gravitational fluid. We identify the radius of the star and, beyond this limit, we apply the necessary conditions so that the de Sitter--Schwarzschild solution
describes the exterior. For the gravitational fluid of the interior solution, we derive the
angular and radial speed of sound as well as the mass of the star. Finally,
in Section \ref{sec5}, we discuss our results and present our conclusions.

\section{Symmetric teleparallel gravity and modifications}
\label{sec2}

In this section, we provide a brief overview of the fundamental aspects of
the Symmetric Teleparallel Equivalent of General Relativity (STEGR) and its
extensions. For a more detailed discussion, we refer the reader to \cite%
{rev1,Heis1}.

In metric-affine gravitational theories the space is characterized by two independent fields. The metric, $g_{\mu \nu }$, which serves as a rule of measuring distances, and the connection, $\Gamma _{~\mu \nu }^{\alpha }$, which is used to define the parallel transport of vectors. The most general connection is analyzed as follows:
\begin{equation}
\Gamma _{~\mu \nu }^{\alpha }=\mathring{\Gamma}_{~\mu \nu }^{\alpha
}+K_{~\mu \nu }^{\alpha }+L_{~\mu \nu }^{\alpha },  \label{generalconnection}
\end{equation}%
\newline
where 
\begin{equation}
\mathring{\Gamma}_{~\mu \nu }^{\alpha }=\frac{1}{2}g^{\kappa \alpha }\left( 
\frac{\partial }{\partial x^{\mu }}g_{\kappa \nu }+\frac{\partial }{\partial
x^{\nu }}g_{\mu \kappa }-\frac{\partial }{\partial x^{\kappa }}g_{\mu \nu
}\right) ,  \label{Christoffel}
\end{equation}%
denote the Christoffel symbols, 
\begin{equation}
K_{~\mu \nu }^{\alpha }=\frac{1}{2}\left( \mathcal{T}_{~\mu \nu }^{\alpha }+%
\mathcal{T}_{\mu ~\nu }^{~a}+\mathcal{T}_{\nu ~\mu }^{~a}\right) ,
\end{equation}%
is the contorsion tensor defined with the help of the torsion, $\mathcal{T}%
_{~\mu \nu }^{\alpha }=2\left( \Gamma _{~\mu \nu }^{\alpha }-\Gamma _{~\nu
\mu }^{\alpha }\right) $, and finally 
\begin{equation}
L_{~\mu \nu }^{\lambda }=\frac{1}{2}g^{\lambda \sigma }\left( Q_{\mu \nu
\sigma }+Q_{\nu \mu \sigma }-Q_{\sigma \mu \nu }\right) ,  \label{disften}
\end{equation}%
is the disformation tensor constructed with the use of the nonmetricity 
\begin{equation}
Q_{\alpha \mu \nu }=\nabla _{\alpha }g_{\mu \nu }.  \label{nonmettensor}
\end{equation}

In metric theories of gravity, like GR, the torsion and the nonmetricity
tensors are zero. As a result, the connection coefficients are given in
terms of the Christoffel symbols and thus the $\Gamma^{\alpha}_{~\mu\nu}$ is
dependent on the metric. In metric-affine theories however, either one of the two - or even both - the torsion and the nonmetricity, are
nonvanishing. In the case of $f(Q)$ gravity, only the nonmetricity is
present. The connection is symmetric, ergo the torsion vanishes, and the spaces is taken to be flat, thus the $\Gamma_{~\mu \nu
}^{\alpha }$ need to be such that the Riemann tensor 
\begin{equation}
R^{\lambda}_{~\mu\nu\kappa} = \frac{\partial}{\partial x^\nu} \Gamma_{~\mu
\kappa }^{\lambda } -\frac{\partial}{\partial x^\kappa} \Gamma_{~\mu \nu
}^{\lambda } + \Gamma_{~\mu \kappa }^{\eta }\Gamma_{~\nu\eta}^{\lambda } -
\Gamma_{~\mu \nu }^{\eta } \Gamma_{~\kappa\eta}^{\lambda }
\end{equation}
is zero. In such a theory the gravitational effects are purely attributed to the nonmetricity.

The fundamental geometric object that enters in the action of the
gravitational theories of this type is the nonmetricity scalar 
\begin{equation}
Q=-g^{\mu \nu }\left( L_{~\beta \mu }^{\alpha }L_{~\nu \alpha }^{\beta
}-L_{~\beta \alpha }^{\alpha }L_{~\mu \nu }^{\beta }\right) .  \label{Qdef}
\end{equation}%
The latter is constructed in such a way, so that the linear in $Q$ action 
\begin{equation}
S_{STEGR}=\int \!\!d^{4}x\sqrt{-g}Q,  \label{STEGR}
\end{equation}%
gives rise to field equations equivalent to those of GR; hence the name
STEGR. This occurs because the nonmetricity $Q$ and the Ricci scalar of GR
(the scalar derived from the curvature tensor constructed with the use of
the Christoffel symbols), differ by a total divergence. In the following section we concentrate in the non-linear modifications of this theory.

\subsection{$f\left( Q\right) $-theory}

As happens in the case of GR and its modifications in terms of different $%
f(R)$ theories, we also obtain modifications to the STEGR by considering
nonlinear functions of the fundamental scalar. We write the extensions
of this theory by invoking the action 
\begin{equation}  \label{fQact}
S=\int \!\!d^{4}x\sqrt{-g} f(Q) + S_m,
\end{equation}
where $f(Q)$ is generally a nonlinear function of $Q$. The $S_m$
denotes the part of the action corresponding to the matter content. For completeness, we
need to mention that the full action includes the curvature $%
R^{\kappa}_{~\lambda\mu\nu}$ and the torsion tensor, which are added with Lagrange
multipliers, so that the $R^{\kappa}_{~\lambda\mu\nu}=0$ and the $\mathcal{T}%
_{~\mu \nu }^{\alpha }=0$ conditions are induced as constraints \cite{fieldeq1,fieldeq2}.

Variation of \eqref{fQact} with respect to the metric yields 
\begin{equation}
\frac{2}{\sqrt{-g}}\nabla _{\lambda }\left( \sqrt{-g}f_Q(Q)P_{\;\mu \nu
}^{\lambda }\right) -\frac{1}{2}f(Q)g_{\mu \nu }+f_Q(Q)\left( P_{\mu \rho
\sigma }Q_{\nu }^{\;\rho \sigma }-2Q_{\rho \sigma \mu }P_{~~~~\nu }^{\rho
\sigma }\right) = \kappa\, T_{\mu\nu},  \label{fieldg}
\end{equation}
where $\kappa=8\pi G/c^4$, 
\begin{equation}
P^\alpha_{\phantom{\alpha}\mu\nu} = -\frac{1}{4} Q^{\alpha}_{\phantom{\alpha}%
\mu\nu} + \frac{1}{2} Q_{(\mu\nu)}^{\phantom{(\mu\nu)}\alpha} + \frac{1}{4}
\left( Q^\alpha - \tilde{Q}^\alpha \right)g_{\mu\nu} - \frac{1}{4}
\delta^{\alpha}_{\phantom{\alpha}(\mu}Q_{\nu)}
\end{equation}
and $Q_\alpha=Q_{\alpha\phantom{\mu}\mu}^{\phantom{\alpha}\mu}$, $\tilde{Q}%
_\alpha = Q^{\mu}_{\phantom{\mu}\alpha\mu}$. The lower index $Q$ appearing
in front of $f$ is used to symbolize the derivative of the function with
respect to $Q$, i.e. $f_Q(Q)= \frac{d f}{dQ}$, $f_{QQ}(Q)=\frac{d^2 f}{dQ^2}$. The $T_{\mu\nu}$ is the energy-momentum tensor associated with the matter
content which we add in the theory. Variation with respect to the (symmetric
and flat) connection, which is an independent field from the metric, leads
to the equations 
\begin{equation}
\nabla_\mu \nabla_\nu \left(\sqrt{-g} f_Q(Q) P^{\mu\nu}_{\phantom{\mu\nu}%
\alpha} \right) =0 ,  \label{fieldgamma}
\end{equation}
assuming of course that the matter does not couple with the connection. Equation \eqref{fieldgamma} is on equal footing with the field equations of the metric \eqref{fieldg}, it also needs to be satisfied. Not any connection for which \eqref{fieldg} form a compatible set of equations will trivialize \eqref{fieldgamma}. 

The field equations \eqref{fieldg} of the metric can also be brought to the form \cite{Zhao} 
\begin{equation} \label{fieldmetE}
 E_{\mu\nu} := f_Q\left( Q\right) G_{\mu \nu }-\frac{1}{2}g_{\mu \nu }\left( f\left(
Q\right) -f_Q\left( Q\right) Q\right) +2f_{QQ}\left( Q\right) P_{~~\mu \nu
}^{\lambda }\nabla _{\lambda }Q= \kappa \, T_{\mu\nu},
\end{equation}
where $G_{\mu \nu }$ is the Einstein tensor of GR and $\kappa=8\pi
G/c^4$. We use the $E_{\mu\nu}$ here to denote the components of the tensor expressing the gravitational part of the theory. A key observation is that, in contrast to what happens in GR, where $\nabla_\mu G^{\mu\nu}\equiv 0$, the divergence $\nabla_\mu E^{\mu\nu}$ is not identically zero. In order to have $\nabla_\mu E^{\mu\nu}=0$ the field equation for the connection, that is Eq. \eqref{fieldgamma}, needs to be satisfied \cite{Zhao}. This is a crucial point for the analysis that follows, because the conservation condition on the matter $\nabla_\mu T^{\mu\nu}=0$, creates a constraint among the functions of the metric, unless Eq. \eqref{fieldgamma} has been satisfied before.

 With the equations expressed in the form of \eqref{fieldmetE} it is easy to see that the
linear version of the theory simply reduces to $G_{\mu\nu} = \kappa_{\text{%
eff}}\; T_{\mu\nu}$. What is more, if $Q=$ const., then the above equations
assume the form 
\begin{equation}
G_{\mu\nu} + \Lambda_{\text{eff}} \; g_{\mu\nu} = \kappa_{\text{eff}} \;
T_{\mu\nu},
\end{equation}
where 
\begin{align}  \label{effLamba}
\Lambda_{\text{eff}} & = \frac{1}{2} \left(Q - \frac{f(Q)}{f_Q(Q)} \right) \\
\kappa_{\text{eff}} & = \frac{\kappa}{f_Q(Q)}
\end{align}
are effective versions of the cosmological and Newton's constants
respectively. Thus, in the $f(Q)\sim Q$ or $Q=$ const. cases the theory is,
at least at the level of the equations, equivalent to GR or GR plus
cosmological constant respectively. Even though the gravitational $f(Q)$-theory is distinct from GR since gravity is now owed to nonmetricity, it is practically indistinguishable - at least at the level of dynamics - from GR. This is true even when considering the motion of matter. For example, a free particle just follows the extremal length curves (geodesics), which are identical irrespectively of whether the theory is based on nonmetricity or curvature. The geodesics just extremize $ds= \sqrt{-g_{\mu\nu}dx^\mu dx^\nu}$, which yields the usual equations involving the Christoffel symbols and not the generic connection.

\section{Static, Spherically Symmetric Spacetimes}
\label{sec3}

\subsection{Relativistic stars within General Relativity}

When seeking interior solutions ($0 \leq r \leq \mathcal{R}$, $\mathcal{R}$ being the radius of the star), the most general form of a static, spherically symmetric geometry in Schwarzschild-like coordinates, $\{ t, r, \theta, \phi \}$, is given by
\begin{equation}
ds^{2}=-e^{A(r)}dt^{2}+e^{B(r)}dr^{2}+r^{2}\left( d\theta ^{2}+\sin
^{2}\theta d\phi ^{2}\right) .  \label{lineelgen}
\end{equation}
with $A(r), B(r)$ being two independent functions of the radial coordinate. In the following we shall be working with the mass function, $m(r)$, which is defined by
\begin{equation}
   e^{-B} = 1 - \frac{2 m}{r}.
\end{equation}

Anisotropic matter is described by a diagonal energy-momentum tensor of the form
\begin{equation}\label{energymomt}
T^{\mu}_{\;\;\nu}= \mathrm{diag}\left(-\rho(r),p_r(r), p_t(r), p_t(r) \right),
\end{equation}
where $p_r,p_t$ are the radial and tangential pressure of matter, respectively, $\rho$ is the energy density of the fluid, while the anisotropic factor is defined by
\begin{equation}
 \Delta = p_t - p_r,
\end{equation}
while in the case of stars made of isotropic matter there is a unique pressure in both directions, $p_r=p_t$, and therefore $\Delta=0$.

In order to find solutions describing hydrostatic equilibrium of compact objects we have to integrate the Tolman-Oppenheimer-Volkoff equations \cite{tolman, OV}
\begin{eqnarray}
    m'(r) & = & 4 \pi r^2 \rho (r)   \\ 
    p_r'(r) & = & - [ \rho(r) + p_r(r) ] \; \frac{A' (r)}{2} + \frac{2 \Delta}{r} \\
    A' (r) & = & 2 \: \frac{m(r) + 4 \pi r^3 p_r(r)}{ r^2 \left( 1 - 2 m(r) / r \right) } 
\end{eqnarray}
where us usual a prime denotes differentiation with respect to $r$.

We integrate the structure equations throughout the star imposing at the center the following conditions 
\begin{equation}
    m(0) = 0, \; \; \; \; \; p_r(0) = p_t(0) = p_{c},
\end{equation}
where $p_{c}$ is the common central pressure. The anisotropic factor vanishes at the center of the star, $\Delta(0)=0$. We impose the following matching conditions at the surface of the star
\begin{equation}
    p_r(\mathcal{R}) = 0, \; \; \; \; m(\mathcal{R}) = M, \; \; \; \; e^{A(\mathcal{R})}=1-2 \frac{M}{\mathcal{R}},
\end{equation}
with $M$ being the stellar mass, taking into account that the exterior vacuum solution ($r > \mathcal{R}$) is given by the Schwarzschild geometry \cite{Schwarzschild:1916uq}
\begin{equation}
    d s^2 = - (1-2M/r) d t^2 + (1-2M/r)^{-1} d r^2 + r^2 (d \theta^2 + \sin^2 \theta d \: \phi^2) .
\end{equation}

Finally, upon numerical integration, once the mass function and the radial pressure are known, the metric potential $A(r)$ is computed by
\begin{equation}
    \displaystyle A (r) = \ln \left( 1 - \frac{2 M}{\mathcal{R}} \right) + 2 \int^r_\mathcal{R} dx \: \frac{m(x) + 4 \pi x^3 p_r(x)}{ x^2 \left( 1 - 2 m(x) / x \right) } .
\end{equation}

\subsection{Exact analytic solution within $f(Q)$ gravity}

The theory is characterized by two basic fields, the metric and the connection. For an excellent review of the geometric aspects of theories with nonmetricity we refer the interested reader to \cite{rev1} and references therein. For the former we adopt the static and spherically symmetric line element as seen in Eq. \eqref{lineelgen}. In what regards the connection, it is well known \cite{Eisenhart} that for a flat
and symmetric connection there can always be found a coordinate system where
its components are zero, i.e. $\Gamma _{~\mu \nu }^{\alpha
}=0$. This is referred to in the literature as the coincident gauge. However,
we need to be careful when we impose such a condition. When assuming a
particular class of metrics, like the one we read from \eqref{lineelgen}%
, we need to consider that we have already fixed the gauge by choosing a
particular coordinate system. The latter may happen to be incompatible with
having $\Gamma _{~\mu \nu }^{\alpha }=0$. 

In these cases, a guide in order to write connections which are compatible
with the field equations, is given by requiring that the connection shares
the symmetries of the metric, i.e. the Killing vectors
\begin{equation} \label{killvecKS}
\begin{split}
& \xi_{1}=\frac{\partial}{\partial\phi}~,~\xi_{2}=\cos\phi \frac{\partial}{\partial\theta} - \sin
\phi \cot{\theta}\frac{\partial}{\partial\phi} \\
& \xi_{3}=\sin\phi\frac{\partial}{\partial\theta}+ \cos\phi \cot{\theta} \frac{\partial}{\partial\phi}~,~ \xi_{4}=\partial_{t}.
\end{split}
\end{equation}
The first three are the generators of spherical symmetry and the fourth imposes staticity. We demand that the above isometries, satisfying $\mathcal{L}_{\xi_i}g_{\mu
\nu }=0$, where $\mathcal{L}$ is the Lie-derivative and $i=1,...,4$, leave invariant the
connection as well
\begin{equation}
   \mathcal{L}_{\xi_i}\Gamma _{~\mu \nu }^{\alpha }=0 , \quad i=1,...,4.
\end{equation} 

The resulting  system of partial differential equations has two general families of solutions \cite{Hohmann2,bh1}. They both play an an important role in the dynamics as they introduce additional degrees of freedom \cite{re11}. 
The first family of solutions, leads to
off-diagonal components in the field equations \eqref{fieldg} for the
metric, which can only be eliminated by
setting $Q=$const. or $f(Q)\sim Q$. Alternatively, one can keep the
off-diagonal terms and consider a non-diagonal energy momentum tensor for
the matter. 

For the second family of connections, the relative non-diagonal
terms can be removed without requiring a constraint on the theory $f(Q)$ or
on $Q$ itself. It is this second connection which we choose in this work.
Its non-zero components are \cite{Hohmann2,bh1,bh2}: 
\begin{equation}
\begin{split}
& \Gamma _{\;tt}^{t}=c_{1}+c_{2}-c_{1}c_{2}\gamma _{1},\quad \Gamma
_{\;tr}^{t}=\frac{c_{2}\gamma _{1}(c_{1}\gamma _{1}-1)}{\gamma _{2}},\quad
\Gamma _{\;rr}^{t}=\frac{\gamma _{1}(1-c_{1}\gamma _{1})(c_{2}\gamma
_{1}+1)-\gamma _{2}\gamma _{1}^{\prime }}{\gamma _{2}^{2}}, \\
& \Gamma _{\;\theta \theta }^{t}=-\gamma _{1},\quad \Gamma _{\phi \phi
}^{t}=-\sin ^{2}\theta \gamma _{1},\quad \Gamma
_{\;tt}^{r}=-c_{1}c_{2}\gamma _{2},\quad \Gamma _{\;tr}^{r}=c_{1}\left(
c_{2}\gamma _{1}+1\right) , \\
& \Gamma _{\;rr}^{r}=\frac{1-c_{1}\gamma _{1}(c_{2}\gamma _{1}+2)-\gamma
_{2}^{\prime }}{\gamma _{2}},\quad \Gamma _{\;\theta \theta }^{r}=-\gamma
_{2},\quad \Gamma _{\;\phi \phi }^{r}=-\gamma _{2}\sin ^{2}\theta ,\quad
\Gamma _{\;t\theta }^{\theta }=c_{1}, \\
& \Gamma _{\;r\theta }^{\theta }=\frac{1-c_{1}\gamma _{1}}{\gamma _{2}}%
,\quad \Gamma _{\phi \phi }^{\theta }=-\sin \theta \cos \theta ,\quad \Gamma
_{t\phi }^{\phi }=c_{1},\quad \Gamma _{\;r\phi }^{\phi }=\frac{1-c_{1}\gamma
_{1}}{\gamma _{2}},\quad \Gamma _{\;\theta \phi }^{\phi }=\cot \theta ,
\end{split}
\label{consol1}
\end{equation}%
where $c_{1}$, $c_{2}$ are constants and $\gamma _{1}(r)$, $\gamma _{2}(r)$
functions of the radial variable $r$. We use the prime to denote derivation
with respect to $r$. 

Nonmetricity theories are based on two basic fields, the metric $g_{\mu\nu}$ and the connection $\Gamma_{\;\mu\nu}^{\kappa}$. The set of components \eqref{consol1} forms the initial ansatz for the connection $\Gamma_{\;\mu\nu}^{\kappa}$ in the same manner line element \eqref{lineelgen} constitutes the ansatz for the metric. For this choice of a connection, and for metric %
\eqref{lineelgen}, the off-diagonal terms of the field equations \eqref{fieldg} are eliminated if we further choose $c_{1}=c_{2}=0$. 

By assuming a matter content leading to an anisotropic energy momentum
tensor of the form \eqref{energymomt}, the remaining independent field equations for the metric read (for
simplicity we drop the argument $Q$ from $f$ and its derivatives): 
\begin{subequations}
\label{feq1}
\begin{align}
& \frac{e^{-B}}{r} \left[ 2 \left(1-e^{\frac{B}{2}}\right) Q^{\prime} f_{QQ} -%
  \frac{1}{r}\left(r B^{\prime}+e^{B}-1\right) f_Q \right] + \frac{1}{2}
  \left(Q f_Q-f\right) = \kappa \, T^0_{~0} , \\
& \frac{1}{2} \left(Q f_Q-f\right) - \frac{e^{-B}}{r^2} \left(e^{B}-r
  A^{\prime}-1\right) f_Q = \kappa \, T^1_{~1} , \\
&  \frac{1}{2} \left(Q f_Q-f\right) + \frac{e^{-B}}{4 r} \left[ 2 r
  A^{\prime\prime} + \left(r A^{\prime} +2\right) \left(A^{\prime} -B^{\prime}
  \right) \right] f_Q - \frac{e^{-B}}{2 r}\left(2 e^{\frac{B}{2}} -r
  A^{\prime}-2\right) Q^{\prime} f_{QQ} = \kappa \, T^2_{~2}.
\end{align}
The equation \eqref{fieldgamma} for the connection reduces to 
\end{subequations}
\begin{equation}  \label{gamma2eq}
\begin{split}
& Q^{\prime} f_{QQ} \left(A^{\prime} \left(r^2-e^{B}
\gamma_2^2\right)-\left(e^{B} \gamma_2^2+r^2\right) B^{\prime}+4
\left(r-e^{B} \gamma_2 \gamma_2^{\prime}\right)\right)+2 \left(r^2-e^{B}
\gamma_2^2\right) Q^{\prime 2} f_{QQQ} \\
& +2 \left(r^2-e^{B} \gamma_2^2\right) Q^{\prime\prime} f_{QQ} =0 ,
\end{split}%
\end{equation}
while the nonmetricity scalar \eqref{Qdef} is expressed as 
\begin{equation}  \label{defQa}
Q = \frac{e^{-B}}{r^2 \gamma_2^2} \left[2 r^2 \gamma_2' - e^{B} \gamma_2^3 \left(A'+B'\right)+r \gamma_2 \left(-r A'+r B'-4\right)+2 \gamma_2^2 \left(r A'-e^{B} \gamma_2'+ e^{B} + 1 \right)\right].
\end{equation}
A key observation here is that the simple choice 
\begin{equation}  \label{gammasol}
\gamma_2 (r) = r e^{-\frac{B}{2}}
\end{equation}
satisfies equation \eqref{gamma2eq} irrespectively of the $f(Q)$ function
and of the $A(r)$ and $B(r)$ appearing in the metric. This connection has been also utilized in the context of scalar nonmetricity to derive black hole solutions \cite{Baha1}. We should underline here that the $\gamma_2$ of \eqref{gammasol} is not the general solution of the connection equation \eqref{gamma2eq}. However, it has the advantage of being the only choice satisfying Eq. \eqref{gamma2eq} irrespectively of the $f(Q)$ function under consideration. Any other solution of \eqref{gamma2eq}, including the most general configuration of \eqref{consol1} with non-zero $c_1$ and $c_2$, would be applicable to a particular $f(Q)$ function. The utilization of such a solution allows for the direct comparison among different $f(Q)$ theories by maintaining a single ansatz for both fundamental fields, i.e. the metric tensor and the connection.

A brief comment is in order at this point. Many works in the literature, attempting to study stellar configurations in $f(Q)$ gravity, either incorporate an over-restricting procedure or include the serious error of disregarding the equation of the connection. In some cases, the coincident gauge is adopted, which however is incompatible with line element \eqref{lineelgen} in spherical coordinates and thus overrestricts the equations; leading only to solutions of the linear $f(Q)\sim Q$ theory, which is however equivalent to GR \cite{coin1,coin2,coin3,coin4,coin5}. In other cases, even though a connection compatible with spherical symmetry is considered, it is usually not checked whether the equation of motion for the connection \eqref{fieldgamma} is satisfied. To make this problem more explicit; a common choice of a connection used in the literature \cite{error1,error2,error3,error4,error5} is the following
\begin{equation} \label{conerror}
  \begin{split}
   & \Gamma^{r}_{\;\theta\theta} = -r, \quad \Gamma^{r}_{\;\phi\phi} = -r \sin^2\theta, \quad \Gamma^{\theta}_{\;r\theta} = \Gamma^{\theta}_{\;\theta r} =\Gamma^{\phi}_{\;r\phi} = \Gamma^{\phi}_{\;\phi r} =\frac{1}{r} , \quad \Gamma^{\theta}_{\;\phi\phi} = -\sin\theta\cos\theta, \\
   & \Gamma^{\phi}_{\;\theta\phi} = \Gamma^{\phi}_{\;\phi \theta} = \cot\theta ,
 \end{split}
\end{equation}
with the rest of components being zero. The resulting nonmetricity scalar is now
\begin{equation} \label{Qerr}
  Q = \frac{e^{-B} \left(1 - e^{B}\right) \left(A^{\prime} + B^{\prime} \right)}{r}.
\end{equation}
It can be seen that connection \eqref{conerror} is a special case of \eqref{consol1} with $c_1=c_2=\gamma_1=0$ and $\gamma_2 (r) = r$. The latter choice for $\gamma_2$ however, unlike \eqref{gammasol}, does not satisfy the field equation for the connection \eqref{gamma2eq}, which in this case yields a restriction between the functions $A$ and $B$ of the metric
\begin{equation}
  \left[\sinh \left(\frac{B}{2}\right) \left(\left(r A^{\prime} + 4\right) Q^{\prime} + 2 r Q^{\prime\prime}\right)+r B^{\prime} \cosh \left(\frac{B}{2}\right) Q^{\prime}\right]  f_{QQ} +2 r \sinh \left(\frac{B}{2}\right) Q^{\prime 2}  f_{QQQ} =0.
\end{equation}
Remember that $Q$ here is given by \eqref{Qerr}, and for a given $f(Q)$ theory this is going to be purely a relation between the functions of the metric, $A$ and $B$. The above equation is trivialized if $Q=$const. or  $f(Q)\sim Q$, which, as is known, describe the same solutions as in the case of GR with or without cosmological constant. Thus, in the generic setting of a non-trivial $f(Q)$ theory, it is not sufficient for the description of the star solution to just solve algebraically \eqref{feq1} with respect to the energy density and the pressures. The constraint set by the field equation of the connection needs to be satisfied. Note that it is not enough to satisfy the analogue of the Tolman-Oppenheimer-Volkoff (TOV) equation. The $\nabla_\mu T^{\mu\nu}=0$ condition on the matter content together with the field equations for the metric \eqref{feq1} are not sufficient conditions. This is because, as we previously mentioned, the covariant derivative of the gravitational tensor $E_{\mu\nu}$ of Eq. \eqref{fieldmetE}, unlike what happens in GR, is not identically zero. The  $\nabla_\mu E^{\mu\nu} =0$ equation holds only if the equation of the connection \eqref{gamma2eq} is satisfied \cite{Zhao}. In order for the latter to happen for a theory dynamically different from GR, we need to find an appropriate $\gamma_2(r)$ function that solves the connection equation. The importance of using a connection compatible with the field equations has also been stressed recently in \cite{Heisenbergnew}.

In our case, the $\gamma_2 (r)$ we derived in \eqref{gammasol} serves to satisfy the field equation for the connection \eqref{gamma2eq}, and thus the constraint in the metric functions is overall avoided. What is more, \eqref{gammasol} satisfies the relevant equation for any  $f(Q)$ theory.  In this respect, the $\gamma_{2,\text{int}} = r e^{-\frac{B}{2}}$ forms an excellent choice for the connection in what regards the interior solution for the star. For the exterior vacuum solution we adopt a different path, which we describe in detail in the next section. It consists of adopting an exterior solution, $\gamma_{2,\text{ext}}$, for which the nonmetricity scalar becomes constant. Thus, we achieve for the inside, a  consistent description of modified gravitational dynamics due to a dynamical nonmetricity, while in the exterior a known vacuum solution compatible with GR plus a cosmological constant. In the same manner that one connects two distinct solutions for the interior (with matter) and for the exterior (vacuum) for the metric, we additionally connect here the two distinct solutions for the other basic field of the theory, the connection.

\subsection{Exterior solution}

We know that the theory becomes dynamically equivalent to GR plus a
cosmological constant when $Q=$const. Hence, the
equations are going to be satisfied by the Schwarzschild-De Sitter metric if, outside the star and in
vacuum, we take the nonmetricity scalar to be constant.
As a result, for the exterior region, we choose to use 
\begin{equation}
ds^2_{\text{ext}} = -\left(1 - \frac{2M}{r} -\frac{\Lambda_{\text{eff}}}{3}
r^2 \right) dt^2 + \frac{dr^2}{\left(1 - \frac{2M}{r} - \frac{\Lambda_{\text{%
eff}}}{3} r^2 \right) } + r^2 \left( d\theta^2 + \sin^2\theta d\phi^2
\right),
\end{equation}
where the effective cosmological constant $\Lambda_{\text{eff}}$ is given by %
\eqref{effLamba}. Note that the resulting Schwarzschild-De Sitter geometry forms the unique type of $f(Q)$ vacuum black-hole solution with reciprocal metric functions $g_{tt}=-1/g_{rr}$ \cite{bh3}.

In order for $Q$ to be constant, we need $\gamma_2$ to satisfy the
differential equation $Q=q$, $q\in \mathbb{R}$, where $Q$ is given by %
\eqref{defQa}. This equation can be easily integrated to give 
\begin{equation}  \label{gammaext}
\begin{split}
\gamma_2 = &  \gamma_{2,\text{ext}} = \frac{1}{12 f_Q} \Bigg[r^3 f - C_1-2 r
\left(q\, r^2-6\right) f_Q \\
& \pm \left[\left(C_1-r^3 f\right)^2 -4 r \left(q\, r^5 f - C_1 q r^2+6
C_1\right)f_Q +4 r \left(72 M+q \, r^3 \left(q\, r^2-6\right)\right)f_Q^2 %
\right]^{\frac{1}{2}}\Bigg],
\end{split}%
\end{equation}
with $C_1$ being an integration constant and $f_Q=f_Q(q)=$const. The
constant $C_1$ plays a crucial role since with this we can continuously
connect the $\gamma_{2,\text{ext}}$ with the $\gamma_{2,\text{int}}$ given
by \eqref{gammasol} at the radius of the star. The equation of motion for the connection now is satisfied due to the fact that the nonmetricity $Q$ is a constant. 

Schwarzschild-(Anti-)De Sitter spacetimes in alternative theories of gravity often arise as exterior black hole solutions with constant curvature; for example in $f(R)$ theories of gravity. In our case it is the constant $Q$ and not the constant $R$ that forms the fundamental scalar of the theory. Note that despite of having a constant nonmetricity outside the star, the interior dynamics is still governed by the non-linear $f(Q)$ terms, which dictate the stellar structure and stability limits in a way distinct from that of Einstein's theory.

\section{The model under consideration}
\label{sec4}

Our following analysis is split into two parts: First, we calculate in a concise manner the exact analytic solution and all the quantities of interest, such as: pressure, energy density, speed of sound, etc. Then, we proceed to investigate what kind of stars this solution represents and how they compare to the ones in the context of GR.

\subsection{The exact interior solution}

Let us consider a theory of the form 
\begin{equation}
f(Q)=Q+\alpha \,Q_{0}\left( \frac{Q}{Q_{0}}\right) ^{\nu },  \label{ftheory}
\end{equation}%
where $\alpha $, $Q_{0}$ and $\nu $ are constants. Even though our main objective is to use this theory to model a basic example as part  of a concise mathematical procedure, this choice is not without physical justification. We already know that the linear theory is dynamically equivalent to GR, and we are also aware that deviations from Einstein's theory cannot be arbitrarily large if we are to satisfy well-established solar system tests. The addition of a polynomial term with a parameter constrained to be small is the simplest modification to this theory. At the same time, by adopting the particular value $\nu=2$ for the exponent (we present for this value the basic graphs in the analysis that follows), the resulting expression consists of the first two terms of a Taylor expansion of an arbitrary $f(Q)$ theory in the region of a small nonmetricity, $Q<<1$. Thus, this choice of $f(Q)$ function can serve to draw conclusions valid as the asymptotic limit of more general theories.

The ansatz we follow for the metric functions, $A(r)$ and $B(r)$ is the following \cite{anss1} 
\begin{align}
A(r)& =C+\left( \frac{r}{r_{1}}\right) ^{2},  \label{ABchoice} \\
B(r)& =\left( \frac{r}{r_{2}}\right) ^{2},
\end{align}%
with $C\in \mathbb{R}$ and $r_{1}$, $r_{2}$ positive constants. The
corresponding geometry describes a singular-free physical solution. Since the solution is based on a metric ansatz without assuming a certain equation-of-state, it should be understood as an effective model rather than a microphysically realistic description.

With these considerations, the resulting energy density and the two
pressures from \eqref{feq1} become 
\begin{equation}  \label{rho}
\begin{split}
\rho = & \frac{Q_0^{-\nu } e^{-\frac{r^2}{r_2^2}}}{2 \kappa r^2 Q^2} \Bigg[4
\alpha (\nu -1) \nu Q_0 r \left(e^{\frac{r^2}{2 r_2^2}}-1\right) Q^{\nu }
Q^{\prime} + 2 Q^2 Q_0^{\nu } \left(\frac{2 r^2}{r_2^2}+e^{\frac{r^2}{r_2^2}%
}-1\right) \\
& + 2 \alpha \nu Q_0 \left(\frac{2 r^2}{r_2^2}+e^{\frac{r^2}{r_2^2}%
}-1\right) Q^{\nu +1} - \alpha (\nu -1) Q_0 r^2 e^{\frac{r^2}{r_2^2}} Q^{\nu
+2}\Bigg],
\end{split}%
\end{equation}
\begin{equation}  \label{pr}
\begin{split}
p_r = \frac{Q_0^{-\nu } e^{-\frac{r^2}{r_2^2}}}{2 \kappa r^2 r_1^2 Q} &  \Bigg[%
\alpha Q_0 Q^{\nu } \left(r_1^2 e^{\frac{r^2}{r_2^2}}
\left((\nu -1) r^2 Q-2 \nu \right)+2 \nu \left(2 r^2+r_1^2\right)\right)\\
& +2 Q Q_0^{\nu } \left(2 r^2-r_1^2 \left(e^{\frac{r^2}{r_2^2}%
}-1\right)\right) \Bigg],
\end{split}%
\end{equation}
\begin{equation}  \label{ptheta}
\begin{split}
p_t = \frac{Q_0^{-\nu } e^{-\frac{r^2}{r_2^2}}}{2 \kappa r r_1^4
r_2^2 Q^2} &  \Bigg[\alpha  Q_0 Q^{\nu } \left((2 (\nu -1) \nu ) r_1^2 r_2^2 Q' \left(r^2-r_1^2 \left(e^{\frac{r^2}{2 r_2^2}}-1\right)\right)+(\nu -1) r r_1^4 r_2^2 Q^2 e^{\frac{r^2}{r_2^2}}\right) \\
& - 2 r Q \left(r^2 \left(r_1^2-r_2^2\right)+r_1^4-2 r_1^2 r_2^2\right) \left(Q Q_0^{\nu }+\alpha  \nu  Q_0 Q^{\nu }\right) \Bigg],
\end{split}%
\end{equation}
respectively. The nonmetricity scalar appearing in the above relations is
calculated from \eqref{defQa}, with the use of \eqref{gammasol} for the
interior, which leads to 
\begin{equation}  \label{Qvalint}
Q(r) = \frac{2 e^{-B}}{r^2} \left(e^{\frac{B}{2}}-1\right) \left(-r
A^{\prime}+e^{\frac{B}{2}}-1\right) = \frac{2 e^{-\frac{r^2}{r_2^2}}}{r^2}
\left(e^{\frac{r^2}{2 r_2^2}}-1\right) \left(e^{\frac{r^2}{2 r_2^2}}-\frac{2
r^2}{r_1^2}-1\right) .
\end{equation}

Using Eq. \eqref{Qvalint} in the expressions \eqref{rho}-\eqref{ptheta}, and
considering theories $f(Q)$ for which $\nu >1$ in \eqref{ftheory}, it is
straightforward to demonstrate that the relevant physical quantities assume
a finite value at $r=0$. The nonmetricity scalar itself becomes zero as $r\rightarrow 0$. In the Taylor expansion, there always exists a
zero-th order term, while the rest that follow are in positive powers of $r$%
. Specifically, for $\nu \geq 2$ the leading $r$-dependent terms are
quadratic: 
\begin{align}
\rho & =\frac{3}{\kappa r_{2}^{2}}+\mathcal{O}(r^{2})  \label{zerpth} \\
p_{r}& =\frac{1}{\kappa }\left( \frac{2}{r_{1}^{2}}-\frac{1}{r_{2}^{2}}%
\right) +\mathcal{O}(r^{2})  \label{zerpth1} \\
p_t & =\frac{1}{\kappa }\left( \frac{2}{r_{1}^{2}}-\frac{1}{r_{2}^{2}}%
\right) +\mathcal{O}(r^{2}).  \label{zerpth2}
\end{align}%
In \eqref{zerpth}-\eqref{zerpth2} we distinguish the finite values at the
origin for the energy density and the two pressures. The pressures are equal
at $r=0$, but differentiate as $r>0$.

Turning to calculate the speed of sound in the radial and tangential directions
from 
\begin{equation}
c_{s,r}^2 =\frac{p_r^{\prime }}{\rho^{\prime }}, \quad c_{s,t}^2 =\frac{p_t^{\prime }}{\rho^{\prime }}
\end{equation}
respectively, we obtain at the limit $r\rightarrow 0$ 
\begin{align}
& \lim\limits_{r\to 0} c_{s,r}^2 = \frac{\left(4
r_2^2-r_1^2\right) \left(Q_0 r_1^2 r_2^2-2 \alpha r_1^2+4 \alpha r_2^2\right)%
}{5 r_1^2 \left(Q_0 r_1^2 r_2^2-2 \alpha r_1^2+8 \alpha r_2^2\right)}, \\
& \lim\limits_{r\to 0} c_{s,t}^2 = \frac{4 \alpha \left(r_1^4-6
r_1^2 r_2^2+8 r_2^4\right)-2 Q_0 r_2^2 \left(r_1^4-3 r_1^2 r_2^2+r_2^4\right)%
}{5 r_1^2 \left(Q_0 r_1^2 r_2^2-2 \alpha r_1^2+8 \alpha r_2^2\right)}
\end{align}
for a theory with $\nu=2$, while for $\nu$ assuming natural number values
with $\nu>2$ we recover 
\begin{align}
& \lim\limits_{r\to 0} c_{s,r}^2 = \frac{4 r_2^2}{5 r_1^2}-\frac{%
1}{5}, \\
& \lim\limits_{r\to 0} c_{s,t}^2 = \frac{2 \left(3 r_1^2
r_2^2-r_1^4-r_2^4\right)}{5 r_1^4}.
\end{align}

\subsection{Physical relevance of the solution}

In what follows, we adopt specific values for the relative parameters for the model, in order to investigate the behavior of the solution and what type of stars it predicts. In each case, we make the relative comparison to GR demonstrating under which different conditions the same metric potentials are produced. We investigate the impact of the coupling $\alpha$, which measures the deviation of GR, on the structural properties of the stars. We also check whether or not well established criteria are met.

\begin{figure}[t]
\centering
\includegraphics[scale=1]{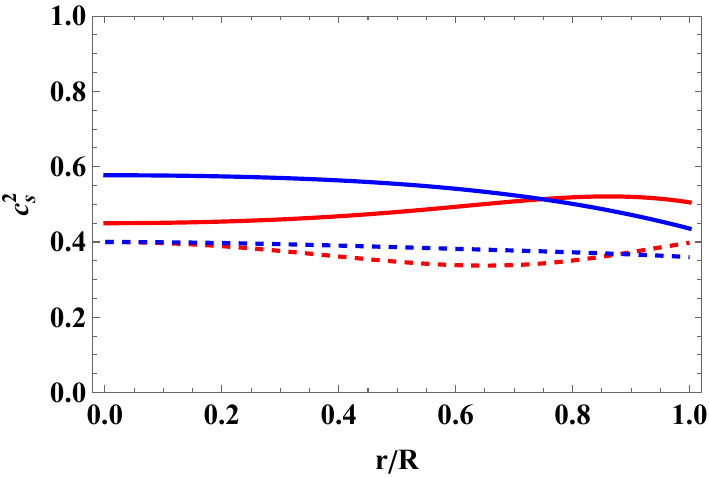}
\caption{The radial (solid lines) and tangential (dashed lines) sound
speeds as functions of the stellar radius considering $\protect%
\nu =2$. The red lines correspond to $\protect\alpha =0.10$ and the blue ones
to $\protect\alpha =0.01$. For the rest of the parameters we have assumed $%
r_{1}=r_{2}=10.00$ km and $Q_{0}=0.01$ km$^{-2}$.}
\label{figsound}
\end{figure}

In Fig. \ref{figsound} we present the plots of the two speeds of sound in
the interior of the star for certain values of the parameters. In particular, we have assumed $r_1=r_2=10 \: km$, $Q_0=1/r_1^2$ and $\alpha=0.01, 0.1$. The values of the parameters $r_1, r_2$ are representative, as they are comparable to the stellar radius.

\begin{figure}[t]
\centering
\includegraphics[scale=1]{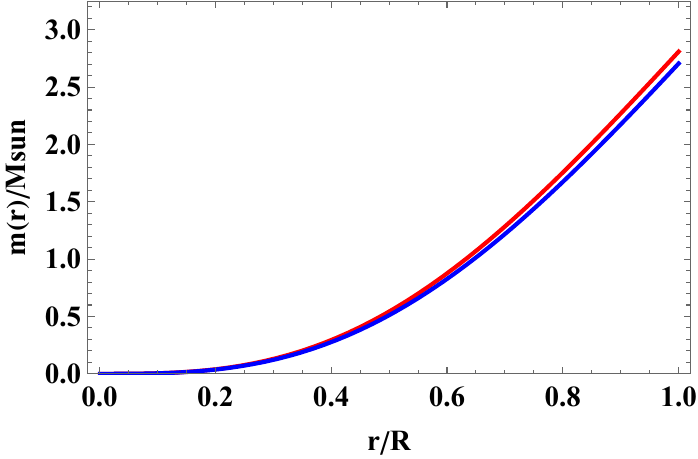}
\caption{The mass function in the interior of the star. The power index
characterizing the theory is $\protect\nu =2$. The red lines correspond to $%
\protect\alpha =0.10$ and the blue ones to $\protect\alpha =0.01$. For both
curves $r_{1}=r_{2}=10.00$ km and $Q_{0}=0.01$ km$^{-2}$.}
\label{figmass}
\end{figure}

The radius of the star is defined as the distance $r=\mathcal{R}$, where $%
p_{r}(\mathcal{R})=0$. As can be seen from \eqref{pr}, this condition
defines a transcendental equation, which can only be solved numerically for
certain values of the parameters. We calculate the mass, $m(r)$, of the star
by parameterizing the $g_{rr}$ component of the metric as 
\begin{equation}
\left( 1-\frac{2m(r)}{r}-\frac{\Lambda _{\text{eff}}}{3}r^{2}\right)
^{-1}=e^{\frac{r^{2}}{r_{2}^{2}}},  \label{masseq}
\end{equation}%
with 
\begin{equation}
\Lambda _{\text{eff}}=\frac{1}{2}\left( Q(\mathcal{R})-\frac{f(Q(\mathcal{R}))}{%
f_{Q}(Q(\mathcal{R}))}\right) .
\end{equation}%
At the radius, $r=\mathcal{R}$, the nonmetricity assumes a value $q=Q(\mathcal{%
R})$, which we consider to remain constant in the outside region $r>r_{s}$.
The resulting mass function for $r\leq \mathcal{R}$ from \eqref{masseq} is 
\begin{equation}
m(r)=\frac{1}{2}r\left( 1-\frac{\Lambda _{\text{eff}}}{3}r^{2}-e^{-\frac{%
r^{2}}{r_{2}^{2}}}\right) .
\end{equation}%
The total mass of the star is of course given by $m(\mathcal{R})=M$. In Fig. %
\ref{figmass} we depict the function of the mass divided by the mass of the
Sun ($M_{\text{Sun}}\sim 1.989\times 10^{30}$ kg) as a function of the
normalized radial distance $r/\mathcal{R}$. As can be seen, stronger deviations from GR, lead to more rapidly rising curves with increasing radius.


\begin{figure}[t]
\centering
\includegraphics[scale=1]{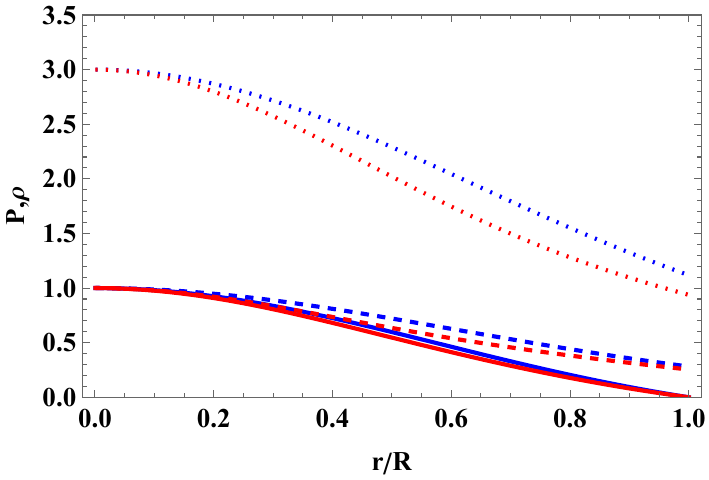}
\caption{
Normalized energy density (dotted curves) and pressures in both directions (radial: solid curves, tangential: dashed curves) versus radial coordinate assuming $\alpha=0.01$ in blue color and $\alpha=0.10$ in red color. All three quantities are positive throughout the star, while at the same time $\rho > p_{r,t}$.
}
\label{figPres}
\end{figure}


\begin{figure}[t]
\centering
\includegraphics[scale=1]{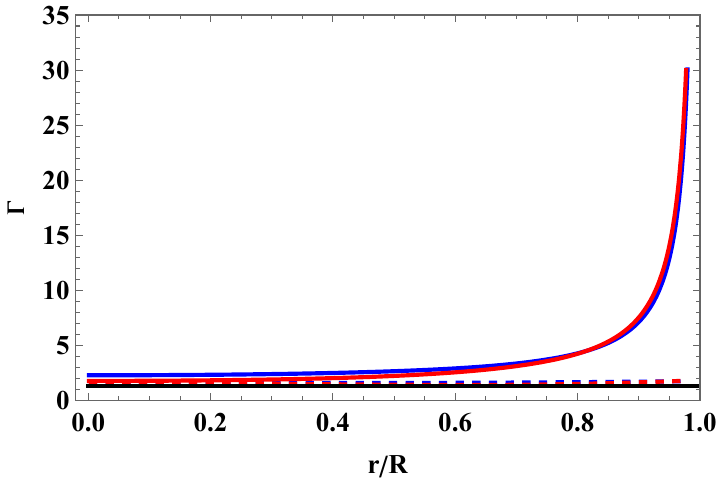}
\caption{
Relativistic adiabatic index, $\Gamma$, in both directions (radial: solid, tangential: dashed) versus radial coordinate considering $\alpha=0.01$ in blue color and $\alpha=0.10$ in red color. The horizontal line in black represents the Newtonian value $4/3$.
}
\label{figGamma}
\end{figure}


For the sake of comparison, we report here the stellar mass and radius as well as the factor of compactness, $C = M/R$, for the cases of GR ($\alpha=0.00$) as well as the two numerical values for $\alpha=0.01, \: 0.10$ for the $f(Q)$ gravity considered here. They are as follows:
\begin{equation}
R=11.21 km, \; \; \; \; M=2.70 M_{\odot}, \; \; \; \; C=M/R=0.358, \; \; \; \; \alpha=0.00
\end{equation}
\begin{equation}
R=11.24 km, \; \; \; \; M=2.71 M_{\odot}, \; \; \; \; C=M/R=0.359, \; \; \; \; \alpha=0.01
\end{equation}
\begin{equation}
R=11.53 km, \; \; \; \; \; M=2.81 M_{\odot}, \; \; \; \; C=M/R=0.363, \; \; \; \; \; \alpha=0.10
\end{equation}

Clearly, an increasing coupling $\alpha$ implies somewhat larger and more massive stars, while at the same time the objects become more and more compact.

Finally, in order to further demonstrate that the solution found here is capable of describing realistic astronomical configurations, we show in Fig. \ref{figPres} and Fig. \ref{figGamma} that the energy conditions \cite{Hawking, Wald} are fulfilled, and that the stability criterion based on the relativistic adiabatic index, $\Gamma > 4/3$, is met, with $4/3$ corresponding to the Newtonian value \cite{Chandra, Moustakidis}. In particular, for compact stars made of anisotropic matter the energy conditions are the following, see e.g. \cite{Panotopoulos:2025lip}:
\begin{equation}
\rho \geq 0, \; \; \; \; \; \rho + p_r + 2 p_t \geq 0, \; \; \; \; \; \rho \pm p_i \geq 0, \; \; \;  i=r,t
\end{equation}
while the relativistic adiabatic index, in both directions, is defined by
\begin{equation}
\Gamma_r = c_{s,r}^2 \left( 1 + \frac{\rho}{p_r} \right), \; \; \; \; \; \Gamma_t = c_{s,t}^2 \left( 1 + \frac{\rho}{p_t} \right).
\end{equation}

It is observed that both $\Gamma_r, \Gamma_t$ remain higher than $4/3$, while at the same time the energy density remains larger than both pressures. Moreover, both pressures and the energy density are positive throughout the star. Therefore, the stability criterion based on $\Gamma$ is met, and all energy conditions are fulfilled. 

The results show that as the coupling $\alpha$ increases, both the energy density and the pressures decrease faster with $r$. Moreover, the impact of $\alpha$ on the relativistic adiabatic index is not considerable. Finally, as far as the sound speeds are concerned, the increase of the parameter $\alpha$ shifts the curves downwards at low $r$, although as we approach the stellar radius the speed of sound becomes larger than the one corresponding to a lower coupling $\alpha$.

Before we conclude our work, a comment is in order here. As we have already mentioned, the primary objective of this study was to establish the mathematical and conceptual consistency of stellar interior solutions in the context of $f(Q)$ gravity, rather than to propose a definitive astrophysical model. What is more, since the exact analytic solution obtained here was based on a metric ansatz without adopting a concrete equation-of-state, it should be understood as an effective model rather than a microphysically realistic description. For those reasons it does not make sense to generate the M-R relationships and compare against current astrophysical constraints. This kind of analysis is performed in order to constrain realistic and well motivated equations-of-state derived using the nuclear physics, particle physics and statistical physics we know. The solution presented here is seen to describe realistic astrophysical configurations such as: neutron stars, quark stars or hyperon stars. The only limitation is the lack of a concrete equation of state, and so a priori we do not know which class of these stars this solution represents.

The deviation of our solution from the one of GR is quantitative as the numerical values of the stellar masses and radii suggest, and this could have considerable impact on other observables such as the frequencies of oscillation of pulsating stars, tidal Love numbers, etc.

\section{Conclusions}
\label{sec5}

In this study, we investigated the structure of static spherically symmetric
stars within the framework of symmetric teleparallel $f\left( Q\right) $%
-gravity. The limit of General Relativity (GR), with or without the
cosmological constant, is recovered in $f\left( Q\right) $-theory when $Q$
is constant. Consequently, if $Q$ is a non-zero constant, the de Sitter–Schwarzschild solution emerges.

We exploited the existence of this known vacuum solution to use it as the exterior metric in our problem. For the interior of the star, we encountered first the connection which is compatible with the field equations induced by a spacetime given by line element \eqref{lineelgen}. It is important to note that this connection can be used in this context with any $f(Q)$ theory. In our investigation we used a power-law $f(Q)$ function (\ref{ftheory}) and, under the ansatz (\ref{ABchoice}), we determined the physical properties of the respective fluid. We determined variables of the parameters, which give a reasonable behavior for the energy density and the two pressures. We identified the radius $r=\mathcal{R}$, where we assume that $Q(\mathcal{R})= \text{const}$. for $r\geq \mathcal{R}$. In this setting we calculated the explicit expression for the mass function. In this context we derived an exact solution to demonstrate the rigor of the procedure. A more physical approach would involve specifying an equation-of-state rather than imposing an ansatz for the metric. However, the resulting solution would then necessarily be numerical. Furthermore, one may go beyond spherical symmetry and add a non-vanishing angular momentum to study properties of rotating stars. We plan to extend this work by investigating astrophysical aspects through the consideration of different equations-of-state in various $f(Q)$ scenarios. Moreover, it would be interesting to study photon orbits and see how they deviate from the null geodesics within GR. We hope to be able to address some of those important topics in the future.

In the model under consideration, we analyzed the radial and tangential sound speeds near the singularity. For values $\nu> 2$ of the power index characterizing the $f\left( Q\right) $-theory,  the variation of the sound speeds is nearly independent of $\nu$. These findings contribute to a deeper understanding of the role of nonmetricity in gravitational models and support the viability of $f(Q)$ gravity in describing compact objects. In a future work we plan to study the definition for the Tolman - Oppenheimer - Volkov formula within the symmetric teleparallel theory of gravity.

\begin{acknowledgments}
We thank the anonymous reviewers for useful comments and suggestions. A.~G was supported by Proyecto Fondecyt Regular 1240247. A.~P thanks the support of Vicerrector\'{\i}a de Investigaci\'{o}n y Desarrollo Tecnol\'{o}gico
(Vridt) at Universidad Cat\'{o}lica del Norte through N\'{u}cleo de
Investigaci\'{o}n Geometr\'{\i}a Diferencial y Aplicaciones, Resoluci\'{o}n
Vridt No - 096/2022 and Resoluci\'{o}n Vridt No - 098/2022. A.~P was partially
supported by Proyecto Fondecyt Regular 2024, Folio 1240514, Etapa 2024. A.~P
thanks the Universidad de La Frontera for the hospitality provided when this
work was carried out.
\end{acknowledgments}


\begin{thebibliography}{99}
%
\bibitem{Teg} M. Tegmark et al., Astrophys. J. 606, 702 (2004)

\bibitem{Kowal} M. Kowalski et al., Astrophys. J. 686, 749 (2008)

\bibitem{Komatsu} E. Komatsu et al., Astrophys. J. Suppl. Ser. 180, 330 (2009)

\bibitem{suzuki11} N. Suzuki et. al., Astrophys. J. 746, 85 (2012)

\bibitem{ade18} Planck Collaboration: Y. Akrami et al. A\&A 641, A10 (2020)

\bibitem{cco1} E. Abdalla et al. JHEAp 34, 49 (2022)

\bibitem{yy} Y. Carloni, O. Luongo and M. Muccino, Phys.\ Rev. D 111, 023512
(2025)

\bibitem{yy1} O. Luongo and M. Muccino, Astron. Astrophys. 641, A174 (2020)

\bibitem{df2} S. Nojiri, S.D. Odintsov and V.K. Oikonomou, Phys. Rept. 692,
1 (2017)

\bibitem{md1} K. Bamba, LHEP 2022, 352 (2022)

\bibitem{md2} S. Shankaranarayanan and J.P. Johnson,\ Gen. Rel.\ Grav. 54,
44 (2022)

\bibitem{hg1} J.D. Barrow, Phys. Lett. B 214, 515 (1988)

\bibitem{hg2} M. Madsen and J.D.\ Barrow, Nucl. Phys. B 323, 242 (1989)

\bibitem{hg3} T. Clifton and J.D. Barrow, Phys. Rev. D 72, 123003 (2003)

\bibitem{hg4} T. Clifton and J.D. Barrow, Class. Quantum Grav. 23, 2351
(2006)

\bibitem{hg5} A.A. Starobinsky, Phys. Lett. B 91, 99 (1980)

\bibitem{fr} T.P. Sotiriou and V. Faraoni, Rev. Mod. Phys. 82, 451 (2010)

\bibitem{oh1} G.\ Papagiannopoulos, S. Basilakos, J.D. Barrow and A.
Paliathanasis, Phys. Rev. D 97, 024026 (2018)

\bibitem{Nester:1998mp} M. Hohmann, Phys. Rev. D 104, 124077 (2021)

\bibitem{Koivisto2} J. B. Jim\'enez, L. Heisenberg and T. S. Koivisto, Phys.
Rev. D 98, 044048 (2018)

\bibitem{Koivisto3} J. B. Jim\'{e}nez, L. Heisenberg, T. S. Koivisto and S.
Pekar, Phys. Rev. D 101, 103507 (2020)

\bibitem{Baha1} S. Bahamonde, J. G. Valcarcel, L. J\"{a}rv and J. Lember,
JCAP 08, 082 (2022)

\bibitem{sc1} L. J\"{a}rv, M. R\"{u}nkla, M. Saal and O. Vilson, Phys. Rev.
D 97, 124025 (2018)

\bibitem{sc2} A.\ Paliathanasis, EPJC 84, 125 (2024)

\bibitem{mf1} A. De, T.-H. Loo and E.N.\ Saridakis, JCAP 03, 050 (2024)

\bibitem{mf2} A. Paliathanasis, Phys. Dark Universe 43, 101388 (2024)

\bibitem{mf4} S. Nojiri and S.D. Odintsov, F(Q) gravity with Gauss-Bonnet
corrections: from early-time inflation to late-time acceleration, (2024)
[arXiv:2406.12558]

\bibitem{rev1} L. Heisenberg, Phys.\ Reports 1066, 1 (2024)

\bibitem{re2} J. Shi, Eur. Phys. J. C 83, 951 (2023)

\bibitem{re1} F. K. Anagnostopoulos, S. Basilakos and E. N. Saridakis, Phys.
Lett. B 822, 136634 (2021)

\bibitem{re3} W. Khyllep, A. Paliathanasis and J. Dutta,\ Phys. Rev. D 103,
103521 (2021)

\bibitem{re4} A. Lymperis, JCAP 11, 018 (2022)

\bibitem{re7} J. Ferreira, T. Barreiro, J.P. Mimoso and N.J. Nunes, Phys.\
Rev. D 108, 063521 (2023)

\bibitem{re8} M. Koussour, N. Myrzakulov, A.H.A. Alfedeel and E.I. Hassan,
Commun. Theor. Phys. 75, 125403 (2023)

\bibitem{re9} A. Paliathanasis, Phys. Dark Univ. 41, 101255 (2023)

\bibitem{re10} A. Paliathanasis, Phys. Dark. Univ. 46, 101585 (2024)

\bibitem{re11} N. Dimakis, A. Paliathanasis, M. Roumeliotis and T.
Christodoulakis, Phys.\ Rev.\ D 106, 043509 (2022)

\bibitem{re12} S. Pradhan, R. Solanki and P.K. Sahoo, JHEAp 43, 258 (2024) \
\ \ \ \ \ \ 

\bibitem{re13} H. Shabani, A. De,\ T.-H. Loo and E.N. Saridakis, EPJC 84,
285 (2024)

\bibitem{re14} G. Subramaniam, A. De, T.-H. Loo and Y.K. Goh, Phys.\ Dark
Univ. 41, 101243 (2023)

\bibitem{re17} S. Nojiri and S.D. Odintsov, Phys. Dark Univ. 45, 101538 (2024)

\bibitem{re18} Y. Carloni and O. Luongo, arXiv:2410.10935 (2024)

\bibitem{re19} S. Capozziello and M. Shokri, Phys. Dark Univ. 37, 101113 (2022)

\bibitem{re20} S. Capozziello and M. Shokri, Phys. Dark Univ. 46, 101698 (2024)

\bibitem{ppr1} D.A. Gomes, J.B. Jimenez, A.J.\ Cano and T.S. Koivisto,
Phys.\ Rev. Lett. 132, 141401 (2024)

\bibitem{ppr2} L. Heisenberg and M. Hohmann, JCAP 03, 063 (2024)

\bibitem{bh1} F. D' Ambrosio, S. D. B. Fell, L. Heisenberg and S. Kuhn,
Phys. Rev. D 105, 024042 (2022)

\bibitem{bh1a} W. Wang, H. Chen and T. Katsuragawa, Phys. Rev. D 105, 024060 (2022)

\bibitem{bh1b} J. Tarciso S.S. Junior and M.E. Rodrigues, Eur. Phys. J. C 83, 475 (2023)

\bibitem{bh2} N. Dimakis, P.A.\ Terzis, A.\ Paliathanasis and T.
Christodoulakis, JHEAp 45, 273 (2025)

\bibitem{bh2a} D. J. Gogoi, A. \"Ovg\"un and M. Koussour, Eur. Phys. J. C 83, 700 (2023)

\bibitem{bh2b} Z.-X. Zhang, C. Lan and Y.-G. Miao, Comparison of Quasinormal Modes of Black Holes in $f(T)$ and $f(Q)$ Gravity, (2025) [arXiv:2501.12800 [gr-qc]]
    
\bibitem{Calza} Marco Calz\'a and Lorenzo Sebastiani, Eur. Phys. J. C 83, 247 (2023)

\bibitem{bh3} S. Bahamonte, J.G. Valcarel, L. J\"{a}rv and J. Lember, JCAP 08,
082 (2022)

\bibitem{Laur1} S. Bahamonde and L. J\"{a}rv, Eur. Phys. J. C 82, 963 (2022)

\bibitem{sol} W. Wang, K. Hu and T. Katsuragawa, Phys. Rev. D 111, 064038 (2025)

\bibitem{Heisenbergnew} L. Heisenberg and C. Pastor-Marcos, Neutron stars in f(Q) gravity, (2025) [2512.03037 [gr-qc]]

\bibitem{oursnew} G. Panotopoulos, A. Lueiza, N. Dimakis and A. Paliathanasis, (2025) [2510.17066 [gr-qc]]

\bibitem{otherpaper} G. Panotopoulos, A. Rinc\'on and I. Lopez, Phys. Dark Univ. 49, 101972 (2025) 

\bibitem{Schwarzschild:1916uq} K.~Schwarzschild,
Sitzungsber. Preuss. Akad. Wiss. Berlin (Math. Phys.) 1916, 189-196 (1916)
[arXiv:physics/9905030 [physics]].

\bibitem{aniso1} R. Ruderman, 
Ann. Rev. Astron. Astrophys., {\bf 10} (1972) 427.

\bibitem{Cadogan:2024mcl} T.~Cadogan and E.~Poisson,
Gen. Rel. Grav. \textbf{56}, no.10, 118 (2024)
[arXiv:2406.03185 [gr-qc]].

\bibitem{Herrera1} L.~Herrera, J.~Martin and J.~Ospino,
J. Math. Phys. \textbf{43}, 4889-4897 (2002)
[arXiv:gr-qc/0207040 [gr-qc]].

\bibitem{Herrera2}
L.~Herrera, J.~Ospino and A.~Di Prisco,
Phys. Rev. D \textbf{77}, 027502 (2008)
doi:10.1103/PhysRevD.77.027502
[arXiv:0712.0713 [gr-qc]].

\bibitem{ExactSol} H. Stephani, D. Kramer, M. Maccallum, C. Hoenselaers and E. Herlt, \textit{Exact Solutions of Einstein’s Field Equations},
Cambridge University Press, Cambridge, England, 2003.

\bibitem{Ovalle:2017fgl} J.~Ovalle,
Phys. Rev. D \textbf{95}, no.10, 104019 (2017)
[arXiv:1704.05899 [gr-qc]].

\bibitem{herrera} L.~Herrera,
Phys. Rev. D \textbf{97}, no.4, 044010 (2018)
[arXiv:1801.08358 [gr-qc]].

\bibitem{karmarkar} K. R. Karmarkar, 
Proc. Ind. Acad. Sci. A {\bf 27}, 56 (1948).

\bibitem{power1} S. Mandal, S. Pradhan, P.K. Sahoo and Tiberiu Harko, Eur. Phys. J. C 83, 1141 (2023) 

\bibitem{power2} D. Mhamdi, F. Bargach, S. Dahmani, A. Bouali and T. Ouali, Phys. Lett. B 859, 139113 (2024) 

\bibitem{Heis1} J. B. Jimenez, L. Heisenberg and T. S. Koivisto, Universe 5, 173 (2019)

\bibitem{fieldeq1} J. B. Jim\'enez, L. Heisenberg and T.S. Koivisto, JCAP 08, 039 (2018)

\bibitem{fieldeq2} M. Hohmann, Universe 7, 114 (2021)

\bibitem{Zhao} D. Zhao, Eur. Phys. J. C 82, 303 (2022)

\bibitem{tolman} R.~C.~Tolman,
Phys. Rev. \textbf{55}, 364-373 (1939).

\bibitem{OV} J.~R.~Oppenheimer and G.~M.~Volkoff,
Phys. Rev. \textbf{55}, 374-381 (1939).

\bibitem{Eisenhart} L. P. Eisenhart, \textquotedblleft Non-Riemannian
Geometry,\textquotedblright\ American Mathematical Society, Colloquium
Publications Vol. VIII, New York, (1927)

\bibitem{Hohmann2} M. Hohmann, Symmetry 12, 53 (2020)

\bibitem{coin1} S.K. Maurya, G. Mustafa, M. Govender and Ksh. N. Singh, JCAP 10 (2022), 003

\bibitem{coin2} A. Ditta, X. Tiecheng, A. Errehymy, G. Mustafa and S. K. Maurya, Eur. Phys. J. C 83, 254 (2023) 

\bibitem{coin3} G. Mustafa, A. Ditta, S. Mumtaz, S.K. Maurya and D. Sofuo\u{g}lu, Chin. J. Phys. 88, 938 (2024)

\bibitem{coin4} M. Zeeshan Gul, S. Rani, M. Adeel and A. Jawad, Eur. Phys. J. C 84, 8 (2024)

\bibitem{coin5} N. Iqbal, S. Khan, M. Alshammari, W. W. Mohammed and M. Ilyas, Eur. Phys. J. C 85, 372 (2025)

\bibitem{error1} R.-H. Lin, X.-H. Zhai, Phys. Rev. D 103, 124001 (2021)

\bibitem{error2} P. Bhar, Eur. Phys. J. C 83, 737 (2023) 

\bibitem{error3} M.A. Alwan, T. Inagaki, B. Mishra and S.A. Narawade, JCAP 09 (2024), 011

\bibitem{error4} P. Bhar, M.R. Shahzad, S. Mandal and P.K. Sahoo, Phys. Dark Univ. 46, (2024) 101686

\bibitem{error5} J.C.N de Araujo and H. G. M. Fortes, (2024) [arXiv:2407.08884 [gr-qc]]

\bibitem{anss1} K.D. Krori and J. Barua, J. Phys. A.: Math. Gen. 8, 508 (1975)


\bibitem{Hawking}
S.~W.~Hawking and G.~F.~R.~Ellis,
``The Large Scale Structure of Space-Time,''
Cambridge University Press, 2011.

\bibitem{Wald}
R.~M.~Wald,
``General Relativity,''
Chicago Univ. Pr., 1984.

\bibitem{Chandra} S.~Chandrasekhar,
Astrophys. J. \textbf{140}, 417-433 (1964)
[erratum: Astrophys. J. \textbf{140}, 1342 (1964)]

\bibitem{Moustakidis} C.~C.~Moustakidis,
Gen. Rel. Grav. \textbf{49}, no.5, 68 (2017)
[arXiv:1612.01726 [gr-qc]].

\bibitem{Panotopoulos:2025lip}
G.~Panotopoulos,
Universe \textbf{11}, 146 (2025)
[arXiv:2504.20347 [gr-qc]].
%
\end{thebibliography}
\end{document}